\documentclass[conference]{IEEEtran}
\pdfoutput=1
\IEEEoverridecommandlockouts
\usepackage{cite}
\usepackage{amsmath,amssymb,amsfonts}
\usepackage{algorithmic}
\usepackage{graphicx}
\usepackage{textcomp}
\usepackage{xcolor}

\def\BibTeX{{\rm B\kern-.05em{\sc i\kern-.025em b}\kern-.08em
    T\kern-.1667em\lower.7ex\hbox{E}\kern-.125emX}}
\begin{document}
\IEEEpubid{\makebox[\columnwidth]{978-1-6654-0144-9/21/\$31.00 \copyright 2021 IEEE \hfill} \hspace{\columnsep}\makebox[\columnwidth]{ }}

\title{Low Resource Species Agnostic Bird Activity Detection \\
}

\author{\IEEEauthorblockN{Mark Anderson}
\IEEEauthorblockA{\textit{SIGMEDIA Group} \\
\textit{Trinity College Dublin}\\
Dublin, Ireland \\
andersm3@tcd.ie}
\and
\IEEEauthorblockN{John Kennedy}
\IEEEauthorblockA{\textit{Department of Mechanical Engineering} \\
\textit{Trinity College Dublin}\\
Dublin, Ireland \\
jkenned5@tcd.ie}
\and
\IEEEauthorblockN{Naomi Harte}
\IEEEauthorblockA{\textit{SIGMEDIA Group} \\
\textit{Trinity College Dublin}\\
Dublin, Ireland \\
nharte@tcd.ie}}

\maketitle

\begin{abstract}
This paper explores low resource classifiers and features for the detection of bird activity, suitable for embedded Automatic Recording Units which are typically deployed for long term remote monitoring of bird populations. Features include low-level spectral parameters, statistical moments on pitch samples, and features derived from amplitude modulation. Performance is evaluated on several lightweight classifiers using the NIPS4Bplus dataset. Our experiments show that random forest classifiers perform best on this task, achieving an accuracy of 0.721 and an F1-Score of 0.604. We compare the results of our system against both a Convolutional Neural Network based detector, and standard MFCC features. Our experiments show that we can achieve equal or better performance in most metrics using features and models with a smaller computational cost and which are suitable for edge deployment.
\end{abstract}


\section{Introduction}
Automatic and remote long-term monitoring of bird populations is of increasing importance for scientific research and conservation as ecological and environmental factors change in the coming years \cite{Johnston20131055, Tilman201773}. Systems that can automatically detect activity and systems that perform species recognition have relevance not only for scientific research and conservation efforts, but as indicators of change in their own right \cite{FIEDLER2009181}. Analysis of audio is highly suitable for monitoring bird populations; many species are identifiable by their vocalisations and visual analysis can be difficult. The vocal mechanisms of songbirds are similar to those of humans, though they exhibit greater control over the amplitude and frequency of their vocalisations \cite{Beckers2011191}. Bird vocalisations vary by species, sub-species and even between different populations \cite{O'ReillyDiff}. Despite this, a species agnostic activity detector which can be further trained on target species is desirable. As the usage of Automatic Recording Units (ARUs) increases \cite{biosciARU}, such a detector operating 24/7 in a remote area of interest, is an essential first step in species identification and population monitoring tasks \cite{RempelARU}.

Automatic analysis of birdsong has followed a similar trajectory to related speech research problems \cite{PriyadarshaniReview}. This is due to the similarity in production between human vocalisations, and bird vocalisations. As a result, the source-filter model is applicable to bird vocalisations \cite{NowickiSF}. Support-Vector Machines (SVMs) \cite{FagerlundSVM, Zhao201799}, Hidden Markov Models (HMMs) \cite{JancovicHMM} and Gaussian Mixture Models (GMMs) \cite{Jancovic2011} have all been employed. Many systems utilise Mel-Frequency Cepstral Coefficients (MFCCs) \cite{PriyadarshaniReview, Dong2020}, and their accompanying deltas and delta-deltas (this is possible due to the applicability of the source-filter model to bird vocalisations). Some systems make additional use of parametric representations such as spectral centroid and roll-off \cite{FagerlundParametric, FagerlundSVM}. 

Recent research into audio-based bird activity detection has used Convolutional Neural Networks (CNNs) and other deep learning methods \cite{GrillCNN, TothCNN, CakirCRNN}. These systems typically make use of spectrograms as features for classification. Such models perform well, but their computational cost is high, when compared with the models and features proposed in this work. If bird activity detection is to be performed on-site, any algorithms must be capable of running in real-time on an embedded device such as an ARU. CNNs are unsuitable in a situation where computational and memory resources are limited. 

This paper proposes computationally less expensive features and methods to perform automatic bird activity detection. The approach removes the requirement for storage of long recordings, flagging recordings of potential interest to researchers and conservationists. Subsequent analysis can then occur off-site to determine specific species or locations, typically with greater computational resources. Work presented here focuses on the activity detection task.
In particular, we introduce features based on amplitude modulation which have been previously deployed by the Institute of Acoustics (IOA) in the long-term monitoring of wind farms \cite{_ioanoise}. There is a relationship between the range of frequencies that a bird species can vocalise and the trill rate, or amplitude modulation, they can achieve in any given call or set of calls \cite{PerformanceLimits}. We also make use of spectral parameters \cite{FagerlundParametric, FagerlundSVM} and pitch based features.

We provide a label for each 1s window of data, which can then be aggregated over an extended period to indicate bird activity within a certain timespan. Several lightweight classifiers using our features are tested on the NIPS4BPlus dataset \cite{MorfiNIPS4BPlus}. Performance compares favourably with alternative approaches using MFCC features and a CNN system based on the system proposed in \cite{GrillCNN}.

\begin{figure*}[ht]
\centering
\includegraphics[width=0.85\textwidth]{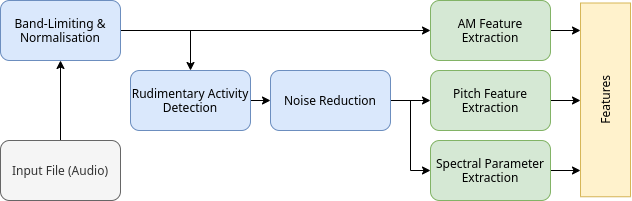}
\caption{\label{fig:programflow}Flowchart of the feature extraction pipeline, with preprocessing steps. Common to all feature extraction algorithms is bandlimiting and normalisation, whereas pitch and spectral features are subject to additional activity detection and noise reduction. Activity detection and noise reduction applied before AM feature extraction will change the envelope of the signal.}
\end{figure*}

\section{Preprocessing and Feature Extraction}
In this section, we present the audio feature extraction techniques used. A flow chart describing the preprocessing and feature extraction pipeline can be seen in Figure \ref{fig:programflow}.

\subsection{Preprocessing}
Field recordings can become contaminated with noise arising from a multitude of sources, including weather conditions, insects, and human activity.  Preprocessing steps undertaken include band-limiting, normalisation, a rudimentary first pass activity detection and noise reduction via filterbanks. The rudimentary activity detection and noise reduction steps are only performed prior to extracting the pitch and spectral parameter based features, as applying them before extracting AM features will contaminate the AM data.

Band limiting is implemented via a bandpass FIR filter. For the experiments presented in this paper, a range of 800Hz - 16KHz was chosen. Normalisation is also performed, due to the varying audio levels in recordings. No calibration for SPL is provided so normalisation consists of scaling the signal such that it spans the entire dynamic range.

Rudimentary activity detection is based upon short-time energy within a specified band compared to the entire signal. If the ratio between the energy in this specified band and the energy contained in the rest of the signal is above a certain threshold (determined experimentally, alongside other preprocessing parameters), that frame is marked as containing activity. The resulting sequence of frames is median filtered with a specified block length, providing a hysteresis effect. The signal is then reconstructed from the frames marked as containing activity.

Noise reduction is achieved using 1/6\textsuperscript{th}-octave filterbanks, centred around 2KHz and covering the same frequency range as the band-limiting filter. In a given frame, 3 out of the 20 bank filters are chosen based upon their normalised energy. The signal is reconstructed from these chosen filterbanks. This technique, while crude, eliminates most of the noise whilst keeping the bird vocalisations intact.

\begin{figure}[t]
\centering
\includegraphics[ trim=2 2 2 2,clip, width=0.45\textwidth]{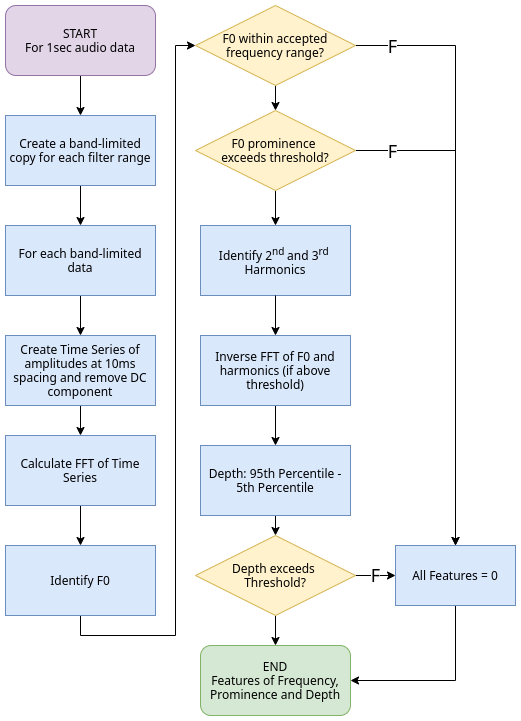}
\caption{\label{fig:am}Flowchart of Amplitude Modulation Feature Extraction, adapted from the IOA method. Whether AM is detected or not is dependent on meeting criteria for 'valid AM frequency', 'AM prominence' and 'AM depth'. These comprise the extracted features used in the classifier.}
\end{figure}

\begin{figure*}[ht]
\centering
\includegraphics[trim=20 20 20 20,clip,width=0.925\textwidth]{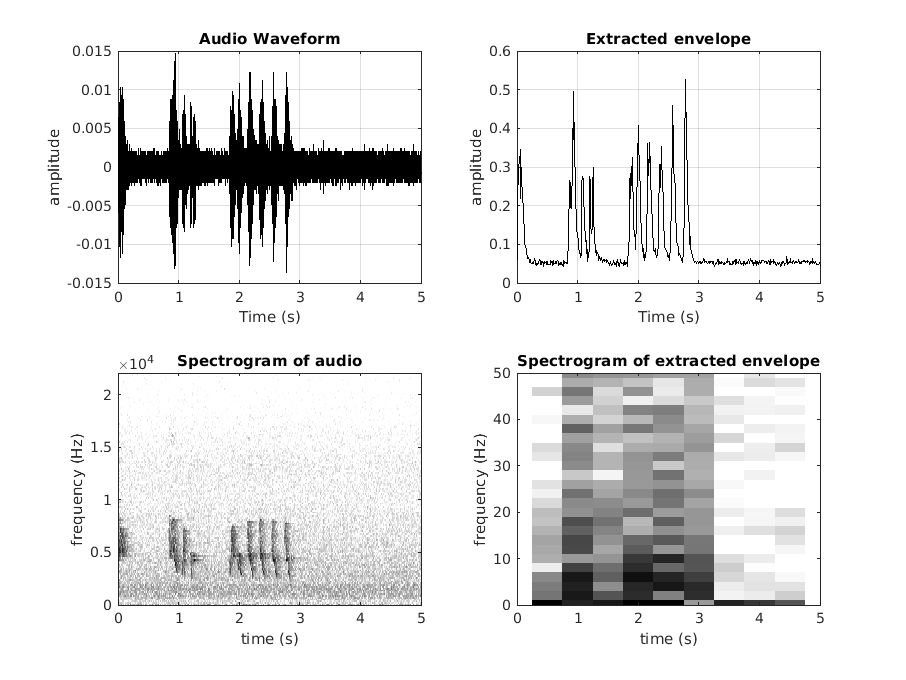}
\caption{\label{fig:spect}Example of Audio file in dataset, and the extracted envelope. Spectrograms of each signal are included. }
\end{figure*}

\subsection{Feature Extraction}
\subsubsection{Pitch Features}
The pitch features extracted are based on the pitch values collected over the 1s analysis period. Within this period, 50 samples of the fundamental frequency are extracted. The pitch extraction algorithm employed is Yin \cite{YIN}. Yin has been shown to be a more accurate pitch detection algorithm for birdsong than other popular methods \cite{O'ReillyYIN}. The features obtained from this section are the statistical moments about the samples taken. We take the first four moments, i.e. Mean, Variance, Skew and Kurtosis. Through these, the shape and distribution of the sampled pitch values can be described, providing an overall description of the pitch within the analysis window. 

Should an analysis window contain no fundamental frequency (i.e. silence or unvoiced sounds), that sample will be represented by a NaN value. NaNs are not included in the calculation of moments. This is done to prevent these values from being lowered due to samples taken by Yin with no pitch.

\subsubsection{Spectral Parameters}
Descriptive, low-level measures taken from the spectral domain are used as input features to the system. These descriptive measures aim to further parameterise the sound within an analysis window. These features are calculated in a manner similar to pitch, in that they are calculated in frames within the analysis window, and the mean and variance values across the entire analysis window are presented as features to the classification system. In our experiments, we utilise spectral centroid and spectral rolloff (taking the sample mean and variance of both). The algorithms used are detailed in \cite{Sharma2020}.

\subsubsection{Amplitude Modulation Features}
The calculation of AM and the features extracted from it takes inspiration from the IOA AM Working Group, whose research centred around detecting amplitude modulation from wind turbines \cite{_ioanoise}. The algorithm has been adapted to suit our application. The primary differences in calculation in AM are the time windows utilised (10ms instead of 100ms) and the bands for which AM is calculated. We process AM based features on four two-octave bands: 500Hz-2KHz, 1KHz-4KHz, 2KHz-8KHz and 4KHz-16KHz. This allows for total coverage of frequencies which the signal had previously been bandlimited to.

The flow chart in Figure \ref{fig:am} outlines the algorithm used to extract AM based features. These are AM Frequency, the prominence of the fundamental frequency of the AM, and the modulation depth of AM as calculated from a reconstructed signal. The information regarding whether or not AM is present in a given window is captured by non-zero values in the other features.

Let us create a sequence $\mathbf{x_{slice}}$ of length $N$ spaced $M$ samples apart as defined in Equation \ref{AM_begin}. 
\begin{align}
    \label{AM_begin}
    \mathbf{x_{slice}} = \{x[0], x[M], ..., x[(N-1)M]\}
\end{align}
This slice of data corresponds to the 1s window on which we are currently calculating features.  In our experiments $N = 100$ and $M = 441$. For each of the two-octave span frequency bands $i$, we create a sequence of energy values. This is the AM envelope sequence for each band $\mathbf{x^i_{env}}$ as in Equation \ref{eq:xienv}:
\begin{align}
\label{eq:xienv}
    x^{i}_{env}[n] = \sqrt{E[\big(\sum_{k=-\infty}^{\infty}x_{slice}[k]h_{i}[k-n]\big)^2]}
\end{align}
The envelope time series is then transformed into a power spectrum (Equation \ref{eq:ps}), and peak detection is performed to find the envelope's fundamental frequency (Equation \ref{AM_freq}). 
\begin{align}
    \label{eq:ps}
    P_{env}^{i}[k] = \frac{1}{N^2}\lvert X^{i}_{env}[k] \rvert^2\\[2ex]
    \label{AM_freq}
    K^{i}_{p} = \arg\!\max_{k} \big(P^{i}_{env}[k]\big)
\end{align}
If this peak is above a prominence threshold, calculated as the ratio of the peak to the expected value of the surrounding bins in Equation \ref{AM_prom} below, it is determined to be the fundamental frequency. 
\begin{align}
    \label{AM_prom}
    Q^i_{peak} = \frac{P^{i}_{env}[K^{i}_{p}]}{\mathbb{E}\big[P^{i}_{env}[K^{i}_{p}-3] + ... + P^{i}_{env}[K^{i}_{p}+3]\big]}
\end{align}
As extraneous noise can be mistaken for AM activity, peak prominence is used to look for pronounced peaks in the frequency domain, indistinct peaks can be classified as noise. If this peak is particularly strong, the second and third harmonics are also checked. Should these harmonics also be deemed prominent by the process detailed above, they are included in the reconstruction of the envelope to calculate depth. The set of the frequency indices used in reconstruction of the signal is $\mathbf{K}^i_{peaks}$.

All valid peaks (i.e. the fundamental, and potentially the second and third harmonics) are kept whilst all the other frequency bins are set to a value of zero. This new power spectrum is transformed back into the time domain as: 
\begin{align}
    \label{AM_recon}
    \mathbf{y^{i}} = \mathbb{R} \big\{ \mathcal{F}^{-1} (X^{i}_{env} [\mathbf{K}^{i}_{peaks}])\big\}
\end{align}
It is from this reconstructed signal that the amplitude modulation depth $D^{i}$ is calculated as:
\begin{align}
    \label{AM_depth_end}
    D^{i} = P_{95}(\mathbf{y^i}) - P_{5}(\mathbf{y^i})
\end{align}
This value is calculated from the 95\textsuperscript{th} and 5\textsuperscript{th} percentile values of the reconstructed signal, as per the IOA method. Thus $D^i$ represents the modulation depth of AM for band $i$. If AM is not detected in an analysis window, a value of zero is assigned to AM Frequency, AM Prominence and AM Depth.

\section{Experimental Setup}
\subsection{Data}
The dataset used in this study is the NIPS4Bplus dataset, a richly annotated birdsong audio dataset \cite{MorfiNIPS4BPlus}. This dataset combines publicly released training data, in the form of bird vocalisations from bird classification challenges \cite{glotin2013neural}, and temporal labels centred around species recognition. These fine grained, temporal annotations make this dataset highly suitable for this work, where the aim is to detect activity in windows of one second. Many other datasets do not contain these temporal annotations. 

The dataset contains 687 recordings sampled at 44.1KHz, comprising 1 hour of audio with vocalisations from 51 species of bird. The dataset also contains recordings where no bird is present (100 files), as well as human and animal activity throughout. Vocalisations span the full range with the majority of energy contained between 1KHz and 10KHz. Recordings took place across 39 locations in France and Spain, starting 30 minutes after sunrise and continuing for a further 3 hours. Recordings were triggered by a $6 dB$ Signal-to-Noise ratio (SNR). The recordings continue until no SNR trigger is detected for 2s. 

As part of this work, the labels have been converted to indicate bird activity in 1s windows with 0.5s overlap. The result of this segmentation is 9 windows per file, covering a 5s audio clip. The labels are a binary classification of whether or not there is bird activity in that window, regardless of vocalisation type or species. The resulting labels are similar to those in datasets such as freefield1010 and warblrb10k (both released by Dan Stowell as part of the DCASE Bird Audio Detection Challenges \cite{StowellBAD2019}). However, the labels of these datasets only provide one label per 10s audio clip. Bird activity is present in 40\% of labels. Our results include a breakdown of selected metrics for both the 'bird present' and 'bird absent' classes.

\subsection{Feature Extraction and Classification}
Feature extraction is performed as outlined in Section 2 and produces an 11x1 feature vector for each analysis window. We denote this feature set as AMPS (Amplitude Modulation, Pitch and Spectrum).  Classifier choices were based on the suitability of an algorithm to embedded applications, i.e. having low memory and computational cost. To this end, we test logistic classification, SVMs and random forest classifiers. In training all models, we have prioritised maximising the F1-Score of the model. Logistic classification is implemented as a base-line classifier. The threshold of the logistic classifier has been tuned to give the highest F1-score. SVMs have been used by \cite{FagerlundSVM}, and our model has been configured similarly. Random forest classifiers have been put to use previously on birdsong in \cite{stowellRF} with good results. Our classifier has been set up with 500 trees, however performance is not significantly reduced with smaller forests meaning further computational saving are possible. We also implement a stacking classifier, an ensemble method combining the logistic, SVM and random forest classifiers. Predictions from each classifier are stacked together and used as input to a final meta-classifier, which is trained through cross-validation. The meta-classifier used is another instance of SVM classification. A stacking method is used due to the heterogeneous nature of the other learners and aims to reduce bias. 

In our comparison with a CNN based detection system, we implement an architecture based on \cite{GrillCNN} using Mel-frequency spectrograms as input.

\begin{table}[ht]
\centering
\caption{\label{hparams}Hyperparameter values for Logistic, SVM and Random Forest classifiers, and parameters used in extraction of amplitude modulation and frequency features}
\begin{tabular}{ll|l}
\hline
\textbf{Classifier} & \textbf{Hyperparameter} & \textbf{Value} \\ \hline
\textbf{Logistic Classifier} & Threshold & 0.45 \\
 & Solver & LBFGS \\ 
\textbf{SVM} & Kernel & RBF \\
 & Regularisation Param. & 1.0 \\
 & Gamma & 0.5 \\ 
\textbf{Random Forest} & Criterion & KL Divergence \\
 & Max. Depth & 8 \\
 & Min. Samples & 8 \\
 & Features per Node & 4 \\
 & N. Trees & 500 \\ \hline
\textbf{Feature} & \textbf{Hyperarameter} & \textbf{Value} \\ \hline
\textbf{AM Features} & Min. Modulation Freq. {[}Hz{]} & 1 \\
 & Max. Modulation Freq. {[}Hz{]} & 10 \\
 & Prominence Cutoff & 3 \\
 & Depth Threshold & 0.01 \\
\textbf{Pitch Features} & Window Length {[}s{]} & 0.02 \\
 & Window Overlap {[}s{]} & 0.01 \\
 & Threshold & 0.3 \\
 \textbf{Spect. Features} & Window Length {[}s{]} & 0.02 \\
 & Window Overlap {[}s{]} & 0.01 \\
\end{tabular}
\end{table}

An 80/20 split between training and test data is performed and results quoted are from this unseen test set. Training is performed using five-fold cross-validation. Classifier hyperparameters were tuned using grid search cross-validation techniques, whereas parameters controlling feature extraction were determined through factorial experiments on the optimised models, coupled with ANOVA testing. Table \ref{hparams} details the hyperparameters of each classifier, and the hyperparameters used for features in the AMPS feature set.

\section{Results}
Table \ref{classifiers_table} reports the performance of the random forest, SVM and a baseline logistic classifier.  Our focus in this work was to prioritise F1-Score, as both precision and recall are valued, however we also wish to evaluate model accuracy, precision and recall separately. To this end, values for each metric are reported when comparing different classifiers and features.

The Logistic Classifier is the best performer in terms of F1-score when using a threshold parameter of $\theta = 0.45$. However, the poor accuracy and precision metrics make it unsuitable for this task. The SVM performs worse in terms of F1, but significantly better in accuracy and precision. Random forest is the best overall performer, with the highest accuracy, precision and only marginal degradation in F1-Score. The stacking classifier shows good accuracy, however, the best overall performer remains the random forest.

\begin{table}[t]
\centering
\caption{\label{classifiers_table}Detection with AMPS features with a variety of classifiers}
\begin{tabular}{l|cccc}
\hline
\textbf{Method} & \multicolumn{1}{l}{\textbf{Acc.}} & \multicolumn{1}{l}{\textbf{F1-Score}} & \multicolumn{1}{l}{\textbf{Prec.}} & \multicolumn{1}{l}{\textbf{Recall}} \\ \hline
Logistic Classifier ($\theta = 0.45$) & 0.538 & 0.611 & 0.462 & 0.900 \\
SVM                 & 0.704 & 0.574 & 0.685 & 0.493 \\
Random Forest       & 0.721 & 0.604 & 0.706 & 0.527 \\
Stacking Classifier & 0.713 & 0.594 & 0.693 & 0.520

\end{tabular}
\end{table}

\begin{table}[t!]
\centering
\caption{\label{class_table}Random forest classification with AMPS features, broken down by 'Bird Absent/Bird Present'}
\begin{tabular}{l|ccc}
\hline
\textbf{Class} & \multicolumn{1}{l}{\textbf{F1-Score}} & \multicolumn{1}{l}{\textbf{Precision}} & \multicolumn{1}{l}{\textbf{Recall}} \\ \hline
Bird Absent         & 0.785 & 0.727 & 0.852  \\
Bird Present        & 0.604 & 0.706 & 0.527  \\ \hline
Weighted Average    & 0.712 & 0.719 & 0.721 
\end{tabular}
\end{table}

\begin{table}[t]
\centering
\caption{\label{features_table}Comparison to state of the art approaches}
\begin{tabular}{c|cccc}
\hline
\textbf{Feature Set} & \textbf{Accuracy} & \textbf{F1-Score} & \textbf{Precision} & \textbf{Recall} \\ \hline
MFCCs                & 0.691                 & 0.561              & 0.655                  & 0.491               \\
CNN                 & 0.674                   & 0.643                   & 0.677                    & 0.612                 \\
AMPS            & 0.721                 & 0.604              & 0.706                  & 0.527                 
\end{tabular}
\end{table}

\begin{table}[t]
\centering
\caption{\label{ops_table}Number of operations for CNN versus random forest system}
\begin{tabular}{c|ccccc}
\hline
\textbf{Classifier} & No. of Ops    \\ \hline
Random Forest                  & 16K CMPs \\
CNN                 &  187k FLOPs      
\end{tabular}
\end{table}

Following this, Table \ref{class_table} details metrics per class for the random forest with AMPS features. Due to imbalance in the dataset, these results allow for a more thorough analysis of the classifier's performance. The higher score of the 'Bird Absent' class is expected due to the higher number of training examples for this class; however the scores for the 'Bird Present' class and the averages of each metric show that our classifier, whilst performing worse on the smaller class, is capable of distinguishing between the two classes. In predictions the classifier incorrectly labeled as 'Bird Absent', the vocalisations were usually 'unvoiced' calls containing little harmonic content, or had significant amounts of noise due to wind, insects, or human activity. Future work must tackle these sources of noise.

Alternate state of the art approaches involve MFCC features and the use of CNNs with spectrograms as input. Therefore in Table \ref{features_table}, we compare the random forest classifier using both our AMPS features and MFCCs as input. The first 13 MFCCs were used, with the algorithm outlined in \cite{Sharma2020}.  The CNN we implemented is based upon work in \cite{GrillCNN}. The AMPS features outperform MFCCs on this task, across all metrics. MFCCs are known to be susceptible to corruption from additive noise, which is prevalent in remote field recordings. The CNN has a higher F1 score, but the performance is not dramatically higher, as might initially be anticipated. Inspection of predictions shows that the CNN performs better on the 'unvoiced' vocalisations and signals with significant levels of in-band noise. Our AMPS features with random forest classification perform better on fainter signals. Both systems fail on recordings with very poor SNR, and recordings with poor dynamic range.

The computational cost of each model is considered in Table \ref{ops_table}. Measurement of CNN complexity in terms of floating point operations required is in line with \cite{cnnCost1}. Furthermore, for edge computing applications the model is pruned and quantised. This was achieved using Tensorflow's built in pruning, quantisation and optimisation functions. The complexity of random forests is measured as the number of comparisons made during classification, using 500 trees, a max depth of eight and four features per node. This gives a worst case figure of 16,000 comparisons, in contrast with 187KFLOPs for the pruned CNN based approach. 

The difference in efficiency between these two operations (CMP vs Floating Point MAC operations) is architecture dependant, however a MAC Operation will be greater than, or equal to, a CMP operation in terms of machine cycles. Therefore our model is at least 10 times less expensive computationally. Thus overall our approach offers a huge computational saving, with a limited sacrifice in performance.

\section{Conclusion}
This paper has proposed an approach to species-agnostic bird activity detection using low resource classifiers alongside AM, spectral and pitch-based features. We have introduced a method of calculating AM based features, to complement features drawn from pitch data and spectral representations. The algorithms used are well within the capabilities of embedded devices, offering performance only slightly below that of a CNN, at a fraction of the computational cost. This approach shows promise as an initial filtering step to aid ornithologists in remote monitoring of bird populations, reducing the amount of data to be stored and processed off-site. With further analysis of the nature of errors made in the system, we plan to tune the feature set and further improve the accuracy of bird activity detection.

\bibliographystyle{IEEEtran}
\bibliography{references}

\begin{thebibliography}{10}
\providecommand{\url}[1]{#1}
\csname url@samestyle\endcsname
\providecommand{\newblock}{\relax}
\providecommand{\bibinfo}[2]{#2}
\providecommand{\BIBentrySTDinterwordspacing}{\spaceskip=0pt\relax}
\providecommand{\BIBentryALTinterwordstretchfactor}{4}
\providecommand{\BIBentryALTinterwordspacing}{\spaceskip=\fontdimen2\font plus
\BIBentryALTinterwordstretchfactor\fontdimen3\font minus
  \fontdimen4\font\relax}
\providecommand{\BIBforeignlanguage}[2]{{%
\expandafter\ifx\csname l@#1\endcsname\relax
\typeout{** WARNING: IEEEtran.bst: No hyphenation pattern has been}%
\typeout{** loaded for the language `#1'. Using the pattern for}%
\typeout{** the default language instead.}%
\else
\language=\csname l@#1\endcsname
\fi
#2}}
\providecommand{\BIBdecl}{\relax}
\BIBdecl

\bibitem{Johnston20131055}
A.~Johnston, M.~Ausden, A.~Dodd, R.~Bradbury, D.~Chamberlain, F.~Jiguet,
  C.~Thomas, A.~Cook, S.~Newson, N.~Ockendon, M.~Rehfisch, S.~Roos, C.~Thaxter,
  A.~Brown, H.~Crick, A.~Douse, R.~McCall, H.~Pontier, D.~Stroud, B.~Cadiou,
  O.~Crowe, B.~Deceuninck, M.~Hornman, and J.~Pearce-Higgins, ``Observed and
  predicted effects of climate change on species abundance in protected
  areas,'' \emph{Nature Climate Change}, vol.~3, no.~12, pp. 1055--1061, 2013.

\bibitem{Tilman201773}
D.~Tilman, M.~Clark, D.~Williams, K.~Kimmel, S.~Polasky, and C.~Packer,
  ``Future threats to biodiversity and pathways to their prevention,''
  \emph{Nature}, vol. 546, no. 7656, pp. 73--81, 2017.

\bibitem{FIEDLER2009181}
W.~Fiedler, ``Chapter 9 - bird ecology as an indicator of climate and global
  change,'' in \emph{Climate Change}, T.~M. Letcher, Ed.\hskip 1em plus 0.5em
  minus 0.4em\relax Elsevier, 2009, pp. 181--195.

\bibitem{Beckers2011191}
G.~Beckers, ``Bird speech perception and vocal production: A comparison with
  humans,'' \emph{Human Biology}, vol.~83, no.~2, pp. 191--212, 2011.

\bibitem{O'ReillyDiff}
C.~O'Reilly, N.~Marples, D.~Kelly, and N.~Harte, ``Quantifying difference in
  vocalizations of bird populations,'' \emph{Proceedings of the Annual
  Conference of the International Speech Communication Association,
  INTERSPEECH}, vol. 2015-January, pp. 3417--3421, 2015.

\bibitem{biosciARU}
L.~S.~M. Sugai, T.~S.~F. Silva, J.~Ribeiro, José~Wagner, and D.~Llusia,
  ``{Terrestrial Passive Acoustic Monitoring: Review and Perspectives},''
  \emph{BioScience}, vol.~69, no.~1, pp. 15--25, 11 2018.

\bibitem{RempelARU}
R.~Rempel, C.~Francis, J.~Robinson, and M.~Campbell, ``Comparison of audio
  recording system performance for detecting and monitoring songbirds,''
  \emph{Journal of Field Ornithology}, vol.~84, no.~1, pp. 86--97, 2013.

\bibitem{PriyadarshaniReview}
N.~Priyadarshani, S.~Marsland, and I.~Castro, ``Automated birdsong recognition
  in complex acoustic environments: a review,'' \emph{Journal of Avian
  Biology}, vol.~49, no.~5, pp. jav--01\,447, 2018.

\bibitem{NowickiSF}
S.~Nowicki, ``Vocal tract resonances in oscine bird sound production: Evidence
  from birdsongs in a helium atmosphere,'' \emph{Nature}, vol. 325, no. 6099,
  pp. 53--55, 1987.

\bibitem{FagerlundSVM}
S.~Fagerlund, ``Bird species recognition using support vector machines,''
  \emph{Eurasip Journal on Advances in Signal Processing}, vol. 2007, 2007.

\bibitem{Zhao201799}
Z.~Zhao, S.-H. Zhang, Z.-Y. Xu, K.~Bellisario, N.-H. Dai, H.~Omrani, and
  B.~Pijanowski, ``Automated bird acoustic event detection and robust species
  classification,'' \emph{Ecological Informatics}, vol.~39, pp. 99--108, 2017.

\bibitem{JancovicHMM}
P.~Jancovic and M.~Kokuer, ``Automatic detection of bird species from audio
  field recordings using hmm-based modelling of frequency tracks,'' \emph{25th
  European Signal Processing Conference, EUSIPCO 2017}, vol. 2017-January, pp.
  1779--1783, 2017.

\bibitem{Jancovic2011}
P.~Jančovič and M.~Koküer, ``Automatic detection and recognition of tonal
  bird sounds in noisy environments,'' \emph{Eurasip Journal on Advances in
  Signal Processing}, vol. 2011, 2011.

\bibitem{Dong2020}
X.~Dong and J.~Jia, ``Advances in automatic bird species recognition from
  environmental audio,'' \emph{Journal of Physics: Conference Series}, vol.
  1544, no.~1, 2020.

\bibitem{FagerlundParametric}
S.~Fagerlund and A.~Härmä, ``Parametrization of inharmonic bird sounds for
  automatic recognition,'' \emph{13th European Signal Processing Conference,
  EUSIPCO 2005}, pp. 1039--1042, 2005.

\bibitem{GrillCNN}
T.~Grill and J.~Schluter, ``Two convolutional neural networks for bird
  detection in audio signals,'' \emph{25th European Signal Processing
  Conference, EUSIPCO 2017}, vol. 2017-January, pp. 1764--1768, 2017.

\bibitem{TothCNN}
B.~Tóth and B.~Czeba, ``Convolutional neural networks for large-scale bird
  song classification in noisy environment,'' \emph{CEUR Workshop Proceedings},
  vol. 1609, pp. 560--568, 2016.

\bibitem{CakirCRNN}
E.~Cakir, S.~Adavanne, G.~Parascandolo, K.~Drossos, and T.~Virtanen,
  ``Convolutional recurrent neural networks for bird audio detection,''
  \emph{25th European Signal Processing Conference, EUSIPCO 2017}, vol.
  2017-January, pp. 1744--1748, 2017.

\bibitem{_ioanoise}
G.~Irvine, M.~Cand, B.~Davis, D.~Coles, S.~Miller, T.~Levet, J.~Shelton,
  J.~Bass, D.~Sexton, and G.~Leventhall, ``Ioa noise working group (wind
  turbine noise), amplitude modulation working group: A method for rating
  amplitude modulation in wind turbine noise,'' 2016.

\bibitem{PerformanceLimits}
J.~Podos and S.~Nowicki, in \emph{Nature's Music: The Science of Birdsong},
  P.~Marler and H.~Slabbekoor, Eds., 2004, pp. 318--342.

\bibitem{MorfiNIPS4BPlus}
V.~Morfi, Y.~Bas, H.~Pamula, H.~Glotin, and D.~Stowell, ``Nips4bplus: A richly
  annotated birdsong audio dataset,'' \emph{PeerJ Computer Science}, vol. 2019,
  no.~10, 2019.

\bibitem{YIN}
A.~De~Cheveigné, ``Yin, a fundamental frequency estimator for speech and
  music,'' \emph{Journal of the Acoustical Society of America}, vol. 111,
  no.~4, pp. 1917--1930, 2002.

\bibitem{O'ReillyYIN}
C.~O'Reilly, N.~Marples, D.~Kelly, and N.~Harte, ``Yin-bird: Improved pitch
  tracking for bird vocalisations,'' \emph{Proceedings of the Annual Conference
  of the International Speech Communication Association, INTERSPEECH}, vol.
  08-12-September-2016, pp. 2641--2645, 2016.

\bibitem{Sharma2020}
G.~Sharma, K.~Umapathy, and S.~Krishnan, ``Trends in audio signal feature
  extraction methods,'' \emph{Applied Acoustics}, vol. 158, 2020.

\bibitem{glotin2013neural}
H.~Glotin, Y.~LeCun, T.~Artieres, S.~Mallat, O.~Tchernichovski, and X.~Halkias,
  ``Neural information processing scaled for bioacoustics, from neurons to big
  data,'' in \emph{Workshop}, 2013.

\bibitem{StowellBAD2019}
D.~Stowell, M.~Wood, H.~Pamuła, Y.~Stylianou, and H.~Glotin, ``Automatic
  acoustic detection of birds through deep learning: The first bird audio
  detection challenge,'' \emph{Methods in Ecology and Evolution}, vol.~10,
  no.~3, pp. 368--380, 2019.

\bibitem{stowellRF}
D.~Stowell and M.~D. Plumbley, ``Automatic large-scale classification of bird
  sounds is strongly improved by unsupervised feature learning,'' \emph{PeerJ},
  vol.~2, p. e488, Jul. 2014.

\bibitem{cnnCost1}
M.~Tan and Q.~Le, ``Efficientnet: Rethinking model scaling for convolutional
  neural networks,'' \emph{36th International Conference on Machine Learning,
  ICML 2019}, vol. 2019-June, pp. 10\,691--10\,700, 2019.

\end{thebibliography}
\end{document}